\documentclass[11pt]{article}
\pdfoutput=1
\title{\textbf{The Distance Precision Matrix: computing networks from nonlinear relationships}}
\date{}
\usepackage{authblk}

\author[1,2]{Mahsa Ghanbari \thanks{mahsa.ghanbari@mdc.berlin.de}}
\author[1,3]{Julia Lasserre \thanks{lasserre@molgen.mpg.de}}
\author[1]{Martin Vingron \thanks{vingron@molgen.mpg.de}}
\affil[1]{Max Planck Institute for Molecular Genetics\\Computational Molecular Biology\\ \small Ihnestrasse 63-73 D-14195 Berlin, Germany} 
\affil[2]{\large Current address: The Berlin Institute for Medical Systems Biology, Max Delbr\"{u}ck Center for Molecular Medicine, \small Robert R\"{o}ssle Str. 10, Berlin 13125, Germany}
\affil[3]{\large Current address: Zalando SE, Berlin, Germany}

\usepackage{float}
\usepackage{multirow}
\usepackage{graphicx}
\usepackage{amsfonts}
\usepackage{amsmath}
\usepackage{algorithm}
\usepackage{algorithmic}
\usepackage{graphicx}
\usepackage{caption}
\usepackage[labelformat = empty,position=top]{subcaption}

\begin{document}

\maketitle

\begin{abstract}
	
\noindent A fundamental method of reconstructing networks, e.g. in the context of gene regulation, relies on the precision matrix (the inverse of the variance-⁠covariance matrix) as an indicator which variables are associated with each other. The precision matrix assumes Gaussian data and its entries are zero for those pairs of variables which are conditionally independent. Here, we propose the Distance Precision Matrix  which is based on a measure of possibly non-⁠linear association, the distance covariance. We provide evidence that the Distance Precision Matrix can successfully compute networks from non-⁠linear data and does so in a very consistent manner across many data situations.
	
\end{abstract}

\section*{Introduction}
Although a somewhat ill-defined task, Network Reconstruction has become ubiquitous in many fields of science. It generally refers to representing associations between variables in the form of a graph, where the nodes correspond to the variables and an edge is drawn whenever an association between the variables is postulated by the chosen network reconstruction method. Examples are Gene Regulatory Networks (GRNs) \cite{Markowetz}, co-authorship networks among researchers \cite{coauthornet}, connectivity networks between brain regions \cite{FMRI}, or social networks among people \cite{Carrington-book}. In machine learning, graphical models have been introduced for this purpose \cite{Bishop, Koller}. 

Typical input data for network reconstruction would be either a similarity matrix or a set of vectors, one for each variable. The latter would be the usual input for GRNs, where the vector contains the gene expression values under many conditions \cite{somogyi}. After centering (and possible standardization to variance 1) call the matrix which contains these vectors as columns $W$. Then $W^t \times W$ is the sample variance-covariance matrix for those variables and constitutes a typical similarity measure among them. In many applications this is the basis for network reconstruction, although other measure (see below) are also in use. 

For a more precise definition of network reconstruction it is helpful to focus not on the edges present in the network, but on the absence of edges. No edge is drawn between two variables when they are deemed to be stochastically independent \cite{Bishop}. This shifts the attention towards determining pairs of variables which are stochastically independent. Under the assumption that data stem from a multivariate Gaussian distribution, i.e. we expect a linear relationship among variables, the correlation coefficient is the appropriate measure to determine independence. It is zero for independent variables, and on sample data there are statistical tests on which one can base a decision whether two variables are deemed independent \cite{Ervin-book}. 

The example of using correlation coefficient as a guide where to draw an edge and where not, highlights another important aspect of network reconstruction. When one variable is correlated with another one, which is in turn correlated with a third variable, one will likely also observe a correlation between the first and the third variable. In the network, however, we would much prefer to see only direct associations represented, as opposed to such inferred ones. This has already been recognized by Fisher, Pearson and Yule (see \cite{Aldrich}) who introduced partial correlation to weed out correlations that can better be explained by a third variable. For gene networks this has been applied, e.g., by de la Fuente \textit{ et al.} \cite{PedroMendes}. Lasserre \textit{ et al.} \cite{Lasserre} used this way of thinking to postulate direct interactions among chromatin modifications and Perner \textit{ et al.} \cite{perner} extended it to associated protein.

In the machine learning literature this has become the basis for Gaussian Graphical Models (GGMs, \cite{Lauritzen, Bishop}). Those rest on the mathematical observation that the entries of the inverse of the variance-covariance matrix are proportional to the full-order partial correlations. This matrix is called the \textit{ precision matrix} and "full order" refers to partial correlation between two variables given the rest of the variables. In practice, based on the assumption that data is multivariate Gaussian, the absence of an edge in a Gaussian Graphical Model corresponds to a very small entry in this so-called precision matrix. It represents not independence of two variables, but conditional independence given the other variables (under the multivariate Gaussian assumption). Conditional independence can be seen as a refinement of the notion of independence when one wants to account for the possibility that two variables might only appear to be linked, whereas in reality it is a third variable the knowledge of which explains the apparent association. Formally, two variables $X$ and $Y$ are conditionally independent with respect to a probability distribution $f$ given a set of variables $Z$, if $f(X,Y | Z)=f(X |Z) f(Y |Z)$ where $f$ is the density function. For Gaussian variables, a partial correlation of 0 between two variables is equivalent to their conditional independence. 

GGMs and the precision matrix are at the core of many network reconstruction methods, even ones that appear to think differently about the problem. Two examples may serve to support this claim. The recently proposed network deconvolution \cite{ND} method proposes to estimate the direct interactions based on an inversion formula in analogy to the summation of a geometric series. A comment on the original paper \cite{commentND} notes the similarity to the precision matrix. Likewise, the Maximum Entropy approach to network reconstruction \cite{Weigtetal,Troyanskaya} has been shown under certain conditions to also correspond closely to the use of the precision matrix (see Appendix in \cite{Morcos-Weigt}).

Network reconstruction, especially when using the precision matrix, rests on a measure of independence or conditional independence among variables. But what when the relationship is not a linear one? Figure 1 shows examples of possible relationships among variables, many of which are non-linear. In practical applications, as in gene regulation, this is a realistic scenario. We look for a quantity that tends to disappear when the two variables are independent, and does so even in the presence of non-linear relationships. 

In principle, mutual information can detect non-linear relationships \cite{Kullback}. It is zero for two independent distributions. However, the sample mutual information is notoriously difficult to estimate \cite{MI1}. In a nutshell, the reason for this is that this requires density estimation, which in itself is an endless topic. In recent years, some progress has been made in this field \cite{MIeq}, and, to name one example, the MIC method \cite{MIC} to identify non-linear relationships tries to solve this binning problem through optimization. 

One alternative measure to mutual information that has been proposed recently by Szekely and coworkers is \textit{ distance covariance} \cite{dcor}. For two random variables, distance covariance is zero exactly when the random variables are statistically independent. The great practical advantage of their definition is that they also provide an estimator. The estimator maps the original data into a high-dimensional space, where the sample distance covariance can be computed as an inner product of these vectors. The estimator does not require any parameter to be selected \cite{commentMIC}.

How can one then detect \textit{ conditional} independence among two variables, given the rest of the variables, when the relationships can be non-linear? There have been attempts to introduce a conditional mutual information \cite{CMI}, which, however, does not alleviate the estimation problems that go along with the use of mutual information. In the realm of biological networks, the work of Califano \cite{ARACNE} has used the information inequality for this purpose. Szekely and Rizzo \cite{pdcor} define partial distance correlation such that Hilbert-space properties are maintained.

The very structure of the estimator for distance covariance makes it particularly suitable for defining a alternative "partial" version of it. Since the estimator computes an inner product in a high-dimensional image of the original data-vectors, it appears natural to proceed in the high-dimensional space as in the linear case and compute partial correlations among the high-dimensional vectors. Not only can one compute partial correlations, one can also merge the high-dimensional vectors into a matrix, which we interpret as the high-dimensional analog of our matrix W. For this, we can compute the variance-covariance matrix and invert it. In this way we combine the best from both worlds: the distance correlation takes care of the non-linear associations, and the precision matrix respects the conditional independence. We will also show that it is generally advantageous to compute this inverse using a regularization method \cite{regpcor}. This approach differs from the one put forward in \cite{pdcor} and we will evaluate both approaches.

The Methods Section of this paper will provide exact definitions of the notions and precisely introduce the \textit{ distance precision matrix} which we use for network construction. For validation it is important to define clearly a simulation setting which challenges the method. This is done in the Results Section, which also reports on results of test runs on simulated and real data from the DREAM challenge. Evaluation is done based on ROC-curves and precision-recall curves (see Methods).

\section{Methods}
\subsection{Distance correlation}
\label{sec:dcor}
Distance correlation has recently been introduced by Szekely and co-workers \cite{dcor} as a measure of association between random variables. It is equal to zero if and only if the random variables are statistically independent. Note that this holds true in general and not only for Gaussian data, which makes the method applicable to the detection of non-linear associations. Here we recapitulate the definition in the specialized form that we will need.

Distance covariance between two univariate random variables $X$ and $Y$, $dcov^2(X,Y)$, is defined as the distance between the joint characteristic function $\phi_{X,Y}$ and the product of its marginal characteristic functions $\phi_{X}$ and $\phi_{X}$ with a particular weight function \cite{dcor}.  The distance correlation is defined as the (non-negative) square root of  distance covariance $dcov^2(X,Y)$ divided by the geometric mean of $dcov^2(X,X)$ and $dcov^2(Y,Y)$. For distributions with finite first moments the distance correlation takes values in $[0,1]$ and the distance correlation is zero if and only if $X$ and $Y$ are independent. In the bivariate normal case $dcov(X,Y)\leq |cov(X,Y)|$, with equality when $|cov(X,Y)|=1$.

The empirical distance correlation for two random variables $X$ and $Y$ with $n$ given samples $X_i, Y_i, i=1,...,n$ is calculated as follows \cite{dcor}. First, distance matrices are defined as  $(a_{ij})= (|X_i-X_j|) $ and  $(b_{ij})= (|Y_i-Y_j|) $. Then the transformed distance matrices $\hat{A}$ and $\hat{B}$, named double centered distance matrices, are obtained from the distance matrices by subtracting the row/column means and adding the grand mean:
\begin{equation}
\hat{A}_{ij}=a_{ij}-\bar{a}_{i.}-\bar{a}_{.j}+\bar{a}
\end{equation}
where $\bar{a}_{i.}=\frac{1}{n} \sum_{k=1}^{n} a_{ik}$,  $\bar{a}_{.j}=\frac{1}{n}\sum_{k=1}^n a_{kj}$ and $\bar{a}=\frac{1}{n^2}\sum_{i,j=1}^n a_{ij}$. The analogous definition is used for $\hat{B}$. 
The sample distance covariance is then defined as the square root of
\begin{equation}
dcov^2_{n}(X,Y)= \frac{1}{n^2} \sum_{i,j=1}^{n} \hat{A}_{ij} \hat{B}_{ij}
\end{equation}
and sample distance correlation $\mathrm{dcor}$ as the square root of 
\begin{equation}
dcor^2_{n}(X,Y)=\frac{dcov^2_{n}(X,Y)}{\sqrt{dcov^2_{n}(X,X)dcov^2_{n}(Y,Y)}}.
\end{equation}

\subsection{Partial distance correlation based on double centered matrices}

Partial distance correlation, in analogy to partial correlation, should be a version of distance correlation which controls for the effect of other variables in the system on the association between two variables. 
Let $\hat{A}$, $\hat{B}$ and $\hat{C}$ be the double centered distance matrices corresponding to variables $X$, $Y$ and $Z$ obtained from $n$ samples. We introduce vectors $V_{\hat{A}}$, $V_{\hat{B}}$ and $V_{\hat{C}}$ as the vector versions of the respective matrices by stringing the columns into a vector. Since the matrices were $n \times n$, the vectors contain $n^2$ elements. We refer to these vectors as double centered vectors and we rewrite sample distance correlation between $X$ and $Y$ as:
\begin{align*}
dcor^2_{n}(X,Y) = cor(V_{\hat{A}},V_{\hat{B}}) 
\end{align*}
Therefore the form of the sample distance correlation offers an approach to defining a notion of a \textit{sample partial distance correlation} by applying partial correlation on the $n^2$-vectors. As in the traditional definition, we regress $V_{\hat{A}}$ and $V_{\hat{B}}$ to $V_{\hat{C}}$ to obtain residuals $r_{\hat{A},\hat{C}}$ and $r_{\hat{B},\hat{C}}$ respectively. Then we define the sample partial distance correlation between $X$ and $Y$ given $Z$ as the correlation between the residuals:
$$pdcor(X,Y;Z)=cor(r_{\hat{A},\hat{C}},r_{\hat{B},\hat{C}})$$
Note that Szekely \& Rizzo \cite{pdcor} define partial distance correlation differently. In the Results Section we will compare the performances of the two approaches. 

\subsection{The Distance Precision Matrix}
Now assume we are given an $n \times p$ matrix $W$, which contains as columns $n$ samples from $p$ random variables. We assume that the columns have been normalized to mean $0$ and standardized to standard deviation $1$. Then $W^t \times W$ is the sample variance-covariance matrix and the precision matrix is defined as its inverse \cite{Bishop}. The entries of the precision matrix $\Lambda$ are related to the full-order partial correlation coefficients by the relationship $pcor(i,j)=-\frac{\Lambda_{ij}}{\sqrt{\Lambda_{ii}\Lambda_{jj}}},\quad i,j=1,...,p,i \neq j$.

We define the \textit{Distance Precision Matrix} by applying the same mechanism in the $n^2$-dimensional space of double centered vectors. For each $X_i$ one computes the double centered matrix and converts it further to a double centered $V$-vector. Let $D$ be the matrix with the double centered $V$-vectors as columns. We define the Distance Precision Matrix (DPM) as the inverse of $D^t \times D$. Edges in a network are drawn for those entries of the Distance Precision Matrix, which are deemed to be non-zero. 

It is an oversimplification to simply speak of inverting the matrices $W^t \times W$ or $D^t \times D$. In many applications one has to deal with the $p>>n$, i.e. the number of variables is much larger than the number of samples. This results in a singular or ill-conditioned matrix $W^t \times W$ or $D^t \times D$.
Even the increased size of the row-space of $D$ ($n^2$) does not necessary alleviate those problems for $D^t \times D$.
Significant efforts in many parts of science, including biology, economics, and finance, have in recent years produced regularization based inversion routines for the variance-covariance matrix (for review see \cite{IlyaPollak}). 

In this study we use the method due to Sch\"afer and Strimmer \cite{regpcor} to estimate and invert a var-covariance matrix. We call the resulting estimators for the inverse the regularized partial correlation (reg-pcor) and the regularized Distance Precision Matrix (reg-DPM), respectively. 

For the purpose of computing a network one needs to decide below which cut-off an edge should be absent. Not following the definition of partial distance correlation from \cite{pdcor} we have no statistical test at hand for computing a cut-off. As long as we compare network construction methods through ROC-curves, we are only interested in the ranking of the edges and the lack of a test for an edge being 0 is not a problem. We will separately study the behavior of the edge weights in order to choose a cut-off.

\subsection{Simulation of data}
The comparison between our method and other, established ones rests to a large degree on simulations. Here we describe the set-up of these simulations. In order to simulate Gaussian data, we first generate a random directed graph $G$ using the R package bnlearn. In the next step, a positive definite matrix from the skeleton of $G$ is generated which then can be used as a covariance matrix for a multivariate Gaussian data. The inverse of this covariance matrix contains zeroes at the missing edges of the given graph G.
With this covariance matrix we can simulate Gaussian data with arbitrary means by using the R package mvtnorm.

\begin{figure}[htb]
\centering
\begin{tabular}{p{.5\textwidth}p{.5\textwidth}}
    \includegraphics[width=.45\textwidth]{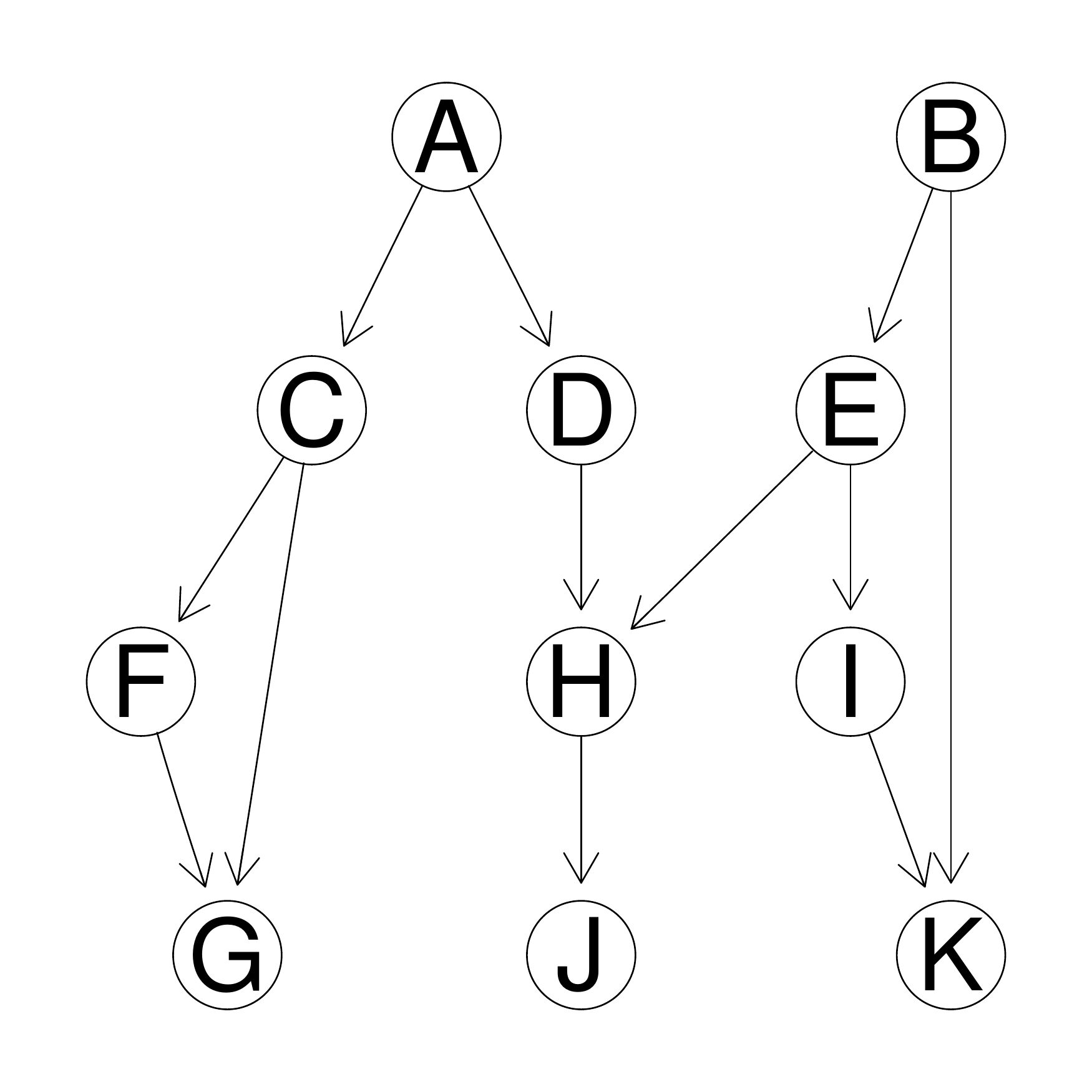}&
    \includegraphics[width=.45\textwidth]{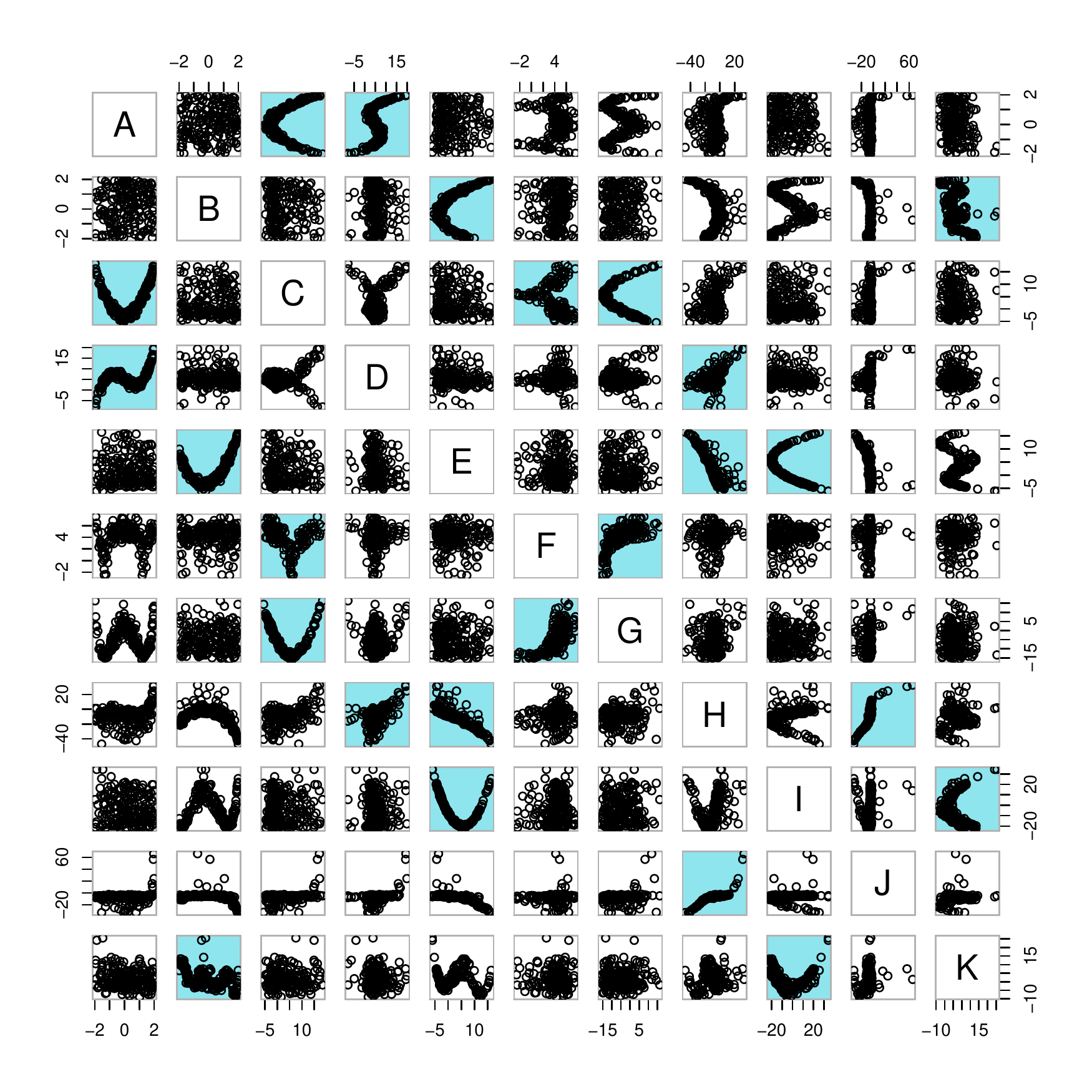}
 \end{tabular}  
\caption{\textbf{Simulation of non-linear data.} The left figure shows the directed graph $GS$ while the left plot shows the scatter plots for one realization of data simulation of the graph GS. The highlighted ones correspond to real edges.}
  \label{fig:gs}
  \end{figure}

For the simulation of nonlinear data we first select a directed graph $GS$ with 11 nodes (see Figure \ref{fig:gs}). We have chosen this graph such that the different arrangements of connections between three nodes are represented: a chain $(x\rightarrow y \rightarrow z),$ a fork $(x\leftarrow y \rightarrow z)$, and a colider $(x \rightarrow y \leftarrow z)$, and a feed-forward loop ($(x\rightarrow y \rightarrow z$ together with $(x \rightarrow z)$). In this graph the value for each node is obtained from its parents using an arbitrarily defined nonlinear function and Gaussian noise is added. Figure \ref{fig:gs} shows the scatter plot of one realization of the simulated data with the direct interactions highlighted in blue. 

\subsection{Data from DREAM challenge}
The DREAM (Dialogue for Reverse Engineering Assessments and Methods) challenge is an annual reverse engineering competition with the aim of fair comparison of network inference methods \cite{Marbach2008,Marbach2012,Prill2010}. The challenge provides synthetic data with non-linear interactions, but also contains biological data-sets with different number and type of variables together with a gold standard network for each data set for the evaluation of the methods. We use the data sets provided by the DREAM challenge, editions DREAM3 and DREAM4 \cite{Marbach2008,Prill2010} as well as the gold standard for evaluation.

\subsection{Competitor methods}
In this study, we compare the performance of DPM and reg-DPM with different other methods. We compare to correlation based methods namely simple Pearson correlation ("cor"), partial correlation ("pcor") and regularized partial correlation ("reg-pcor"). For the latter we use the method introduced by Sch\"{a}fer and Strimmer \cite{regpcor}. To specifically address the ability to recognize direct interactions we compare the performance of DPM and reg-DPM with distance correlation ("dcor"). Furthermore,  we compare to the Szkely et al. \cite{pdcor} version of partial distance correlation ("pdcor"). In terms of well-known network reconstruction methods we compare to the ARACNE \cite{ARACNE} "arac", which is based on mutual information to recognize non-linear relationships, and to network deconvolution \cite{ND} ("nd") to which we supply the variance-covariance matrix as input. The latter was developed particularly to detect direct interactions, although it has been observed that the principle is similar to the inversion of the variance-covariance matrix \cite{commentND}.

\subsection{Evaluation methodology}
We compare methods using the receiver-operating characteristic (ROC) curve  as well as the area under the ROC curve noted AUROC, together with the precision recall (PR) curve  and the area under it denoted AUPRC. In network reconstruction one generally expects only a small fraction of the possible edges among nodes to be correct. In such imbalanced situations it has been shown that the PR curve is very informative in order to distinguish method performances \cite{PR}. Edges were sorted in ascending order of the absolute value of their scores given by the respective method and the ROC and PR curves were built on growing the network starting from the highest score down to the lowest one.

The practical utility of ROC and PR curves is limited in so far as  it may remain hard for a concrete application to decide on a cut-off until which edges are accepted. In the classical setting of Gaussian data, statistical test are available which, together with multiple testing correction, provide guidance. However, for many other measure, including our DPM and reg-DPM, no statistical test are available (yet). In principle, permutation test can be applied \cite{dcor, pdcor} but this may be very compute-intensive. To elucidate this issue we will provide and discuss smoothed histograms of edge-scores.

\section{Results}
\subsection{Performance comparison on Gaussian data} 
\begin{figure}[htb]
\centering
\begin{tabular}{p{.5\textwidth}p{.5\textwidth}}
    \includegraphics[width=.45\textwidth]{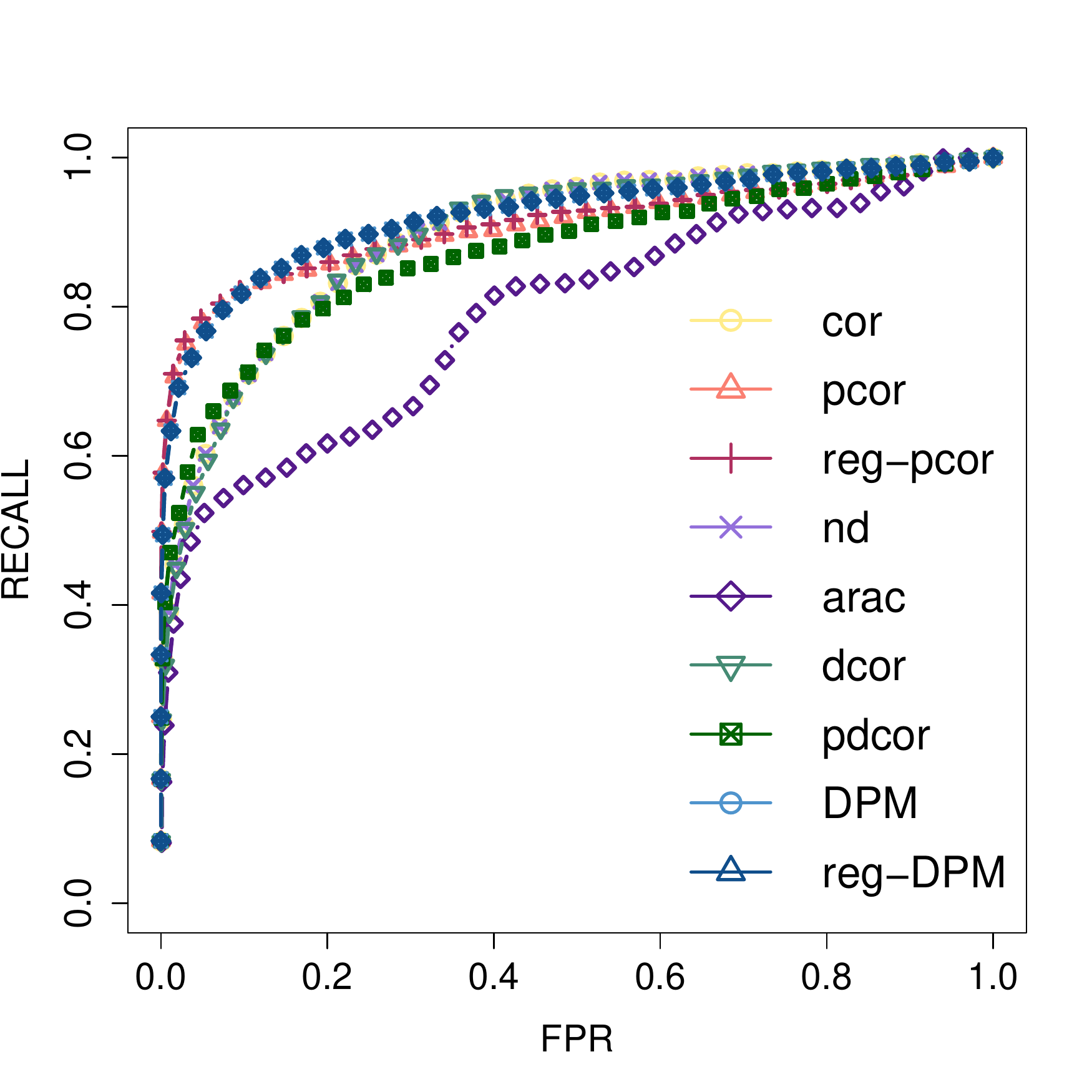}&
    \includegraphics[width=.45\textwidth]{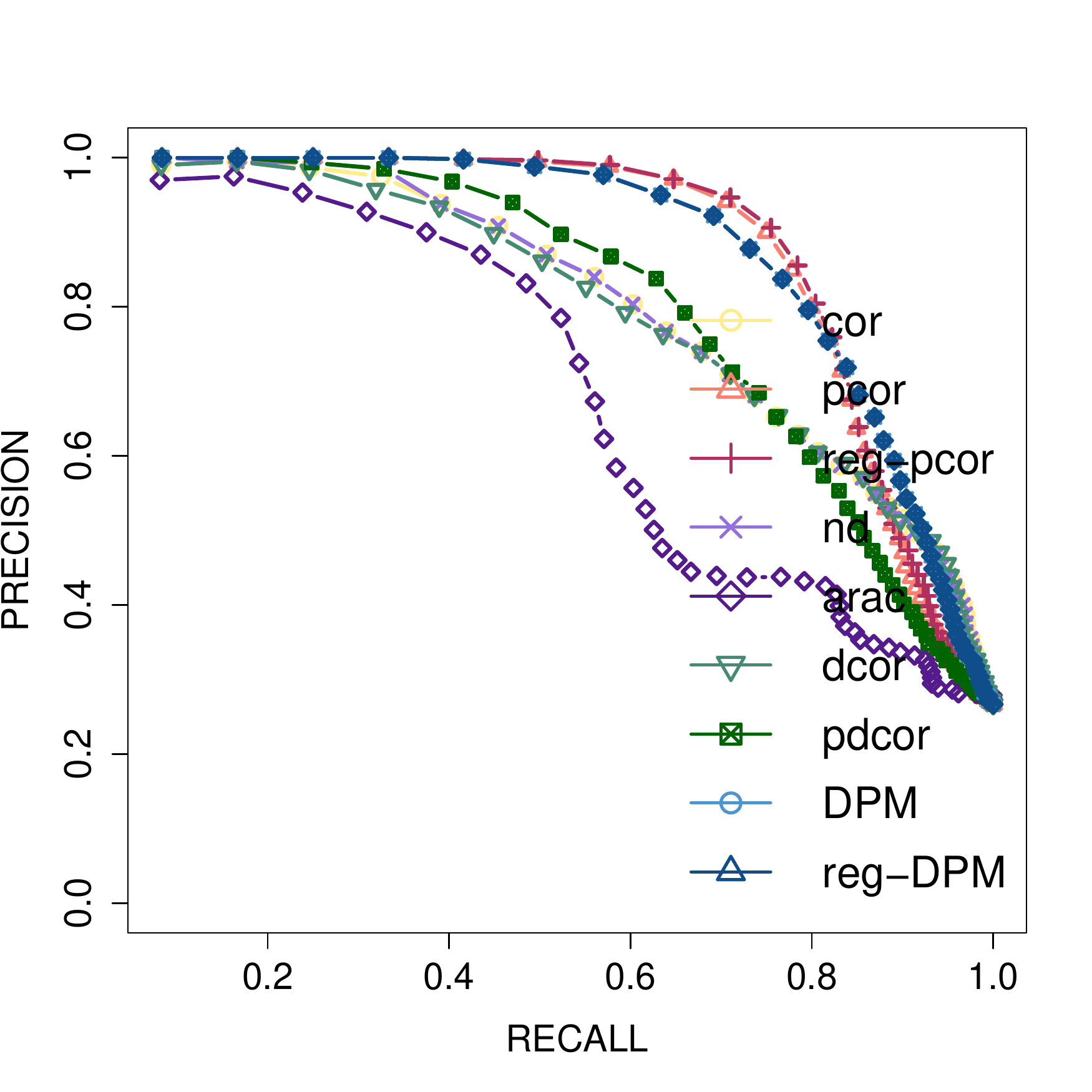}\\
     \includegraphics[width=.45\textwidth]{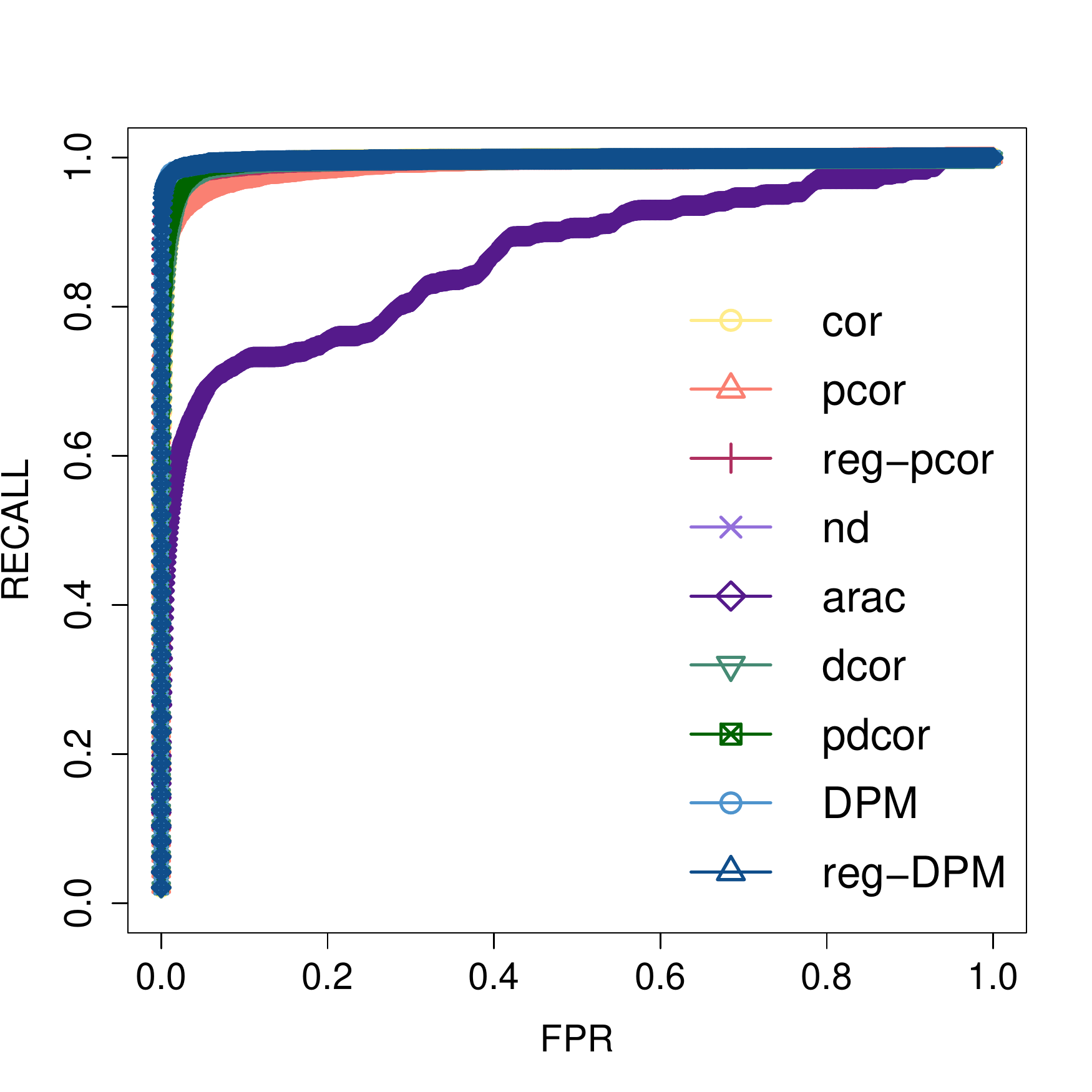}&
    \includegraphics[width=.45\textwidth]{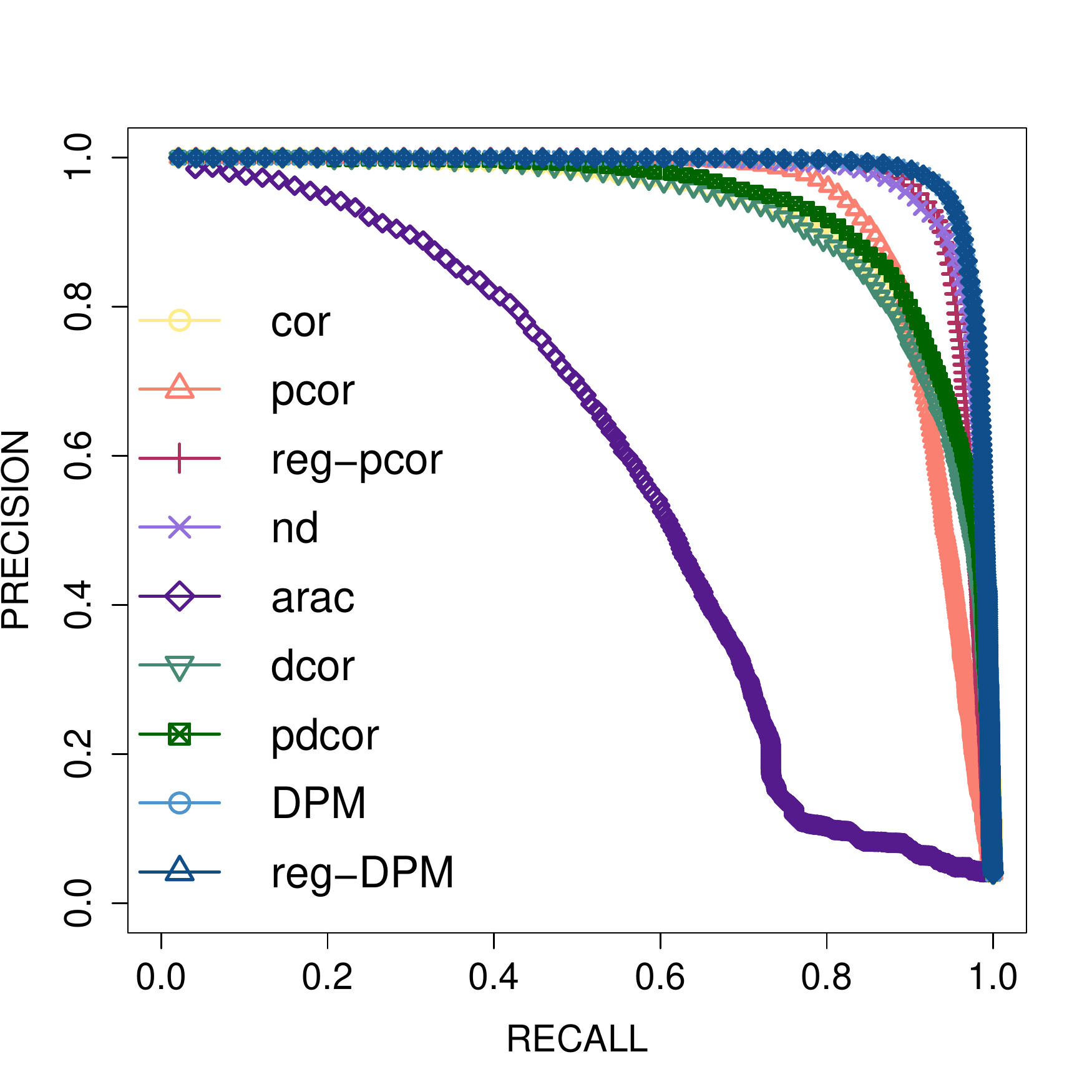}\\
    
 \end{tabular}  
\caption{\textbf{Performance on simulated Gaussian data.} The left subplots show ROC curves, while the right subplots show PR curves. The first row for a network with 10 nodes and second row for a network with 50 nodes.}
  \label{fig:gcomp}
  \end{figure}

In the case of multivariate Gaussian data, the performance of partial distance correlation should be as good as partial correlation in order to be a proper substitution for partial correlation. This has in principle already been found to hold true for distance correlation \cite{dcor, commentMIC}.
We simulate 200 samples of Gaussian data from networks with 50 nodes and additionally add standard Gaussian noise.  This was repeated 100 times and averages are reported.
Figure \ref{fig:gcomp} shows the ROC and PR curves.

The lower curve is ARACNE, which is particularly geared towards non-linear data rather than the Gaussian data in this example.
As for the other methods, they perform almost identically in the ROC curve. The PR curve shows differences between methods, where in particular regularized partial correlation, network deconvolution, and the Distance Precision Matrix as well as its regularized version perform well. Thus, DPM does not lose out over the traditional precision matrix, which is specifically geared towards the Gaussian data. In this "easy" case with many more samples than nodes, regularization visibly improves the traditional precision matrix, but for DPM both the naive inversion and the regularized inversion work well.

\subsection{Performance comparison on nonlinear relationships} 
\begin{figure}[htb]
\begin{tabular}{p{.5\textwidth}p{.5\textwidth}}
\centerline{\includegraphics[width=0.5\textwidth]{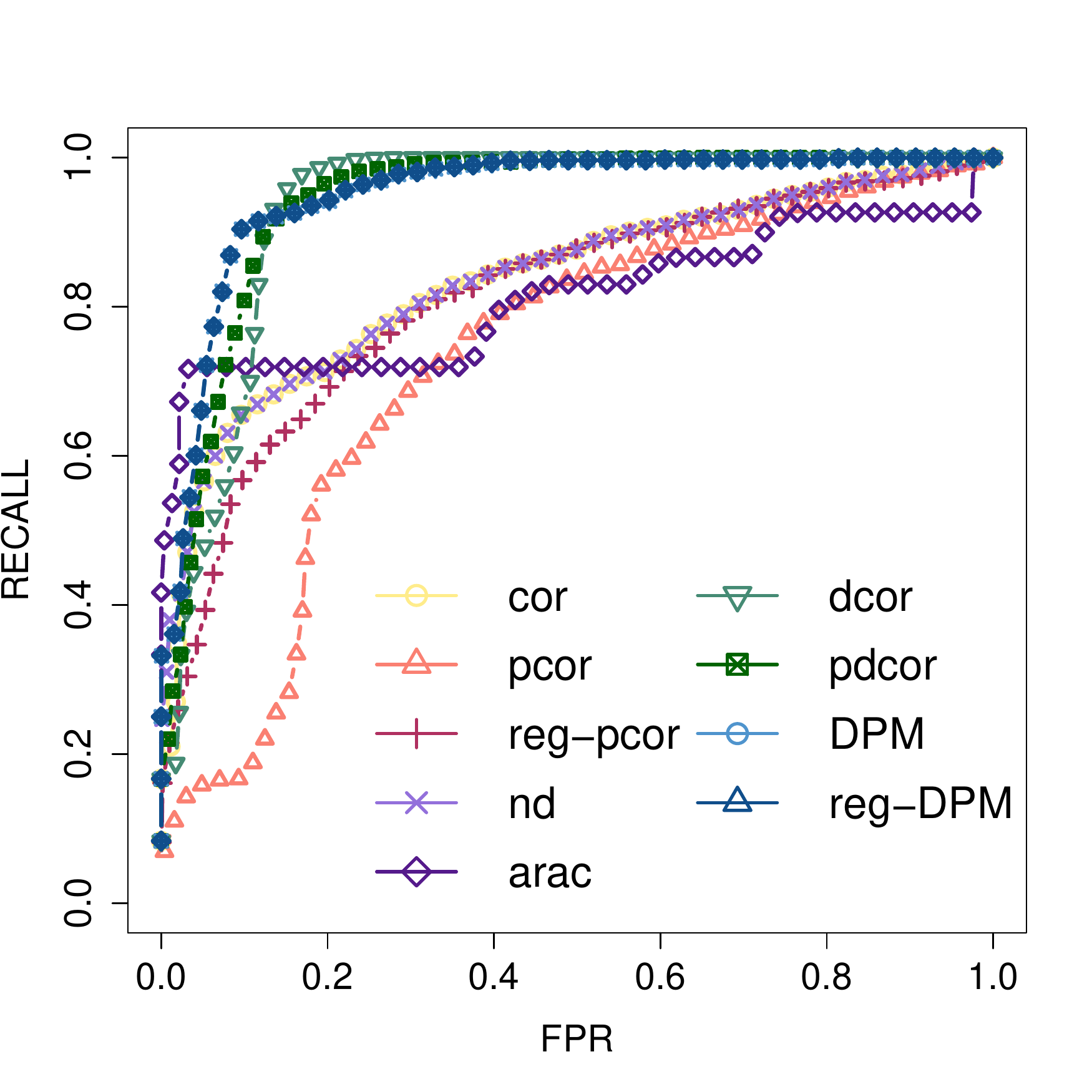}}&
\centerline{\includegraphics[width=0.5\textwidth]{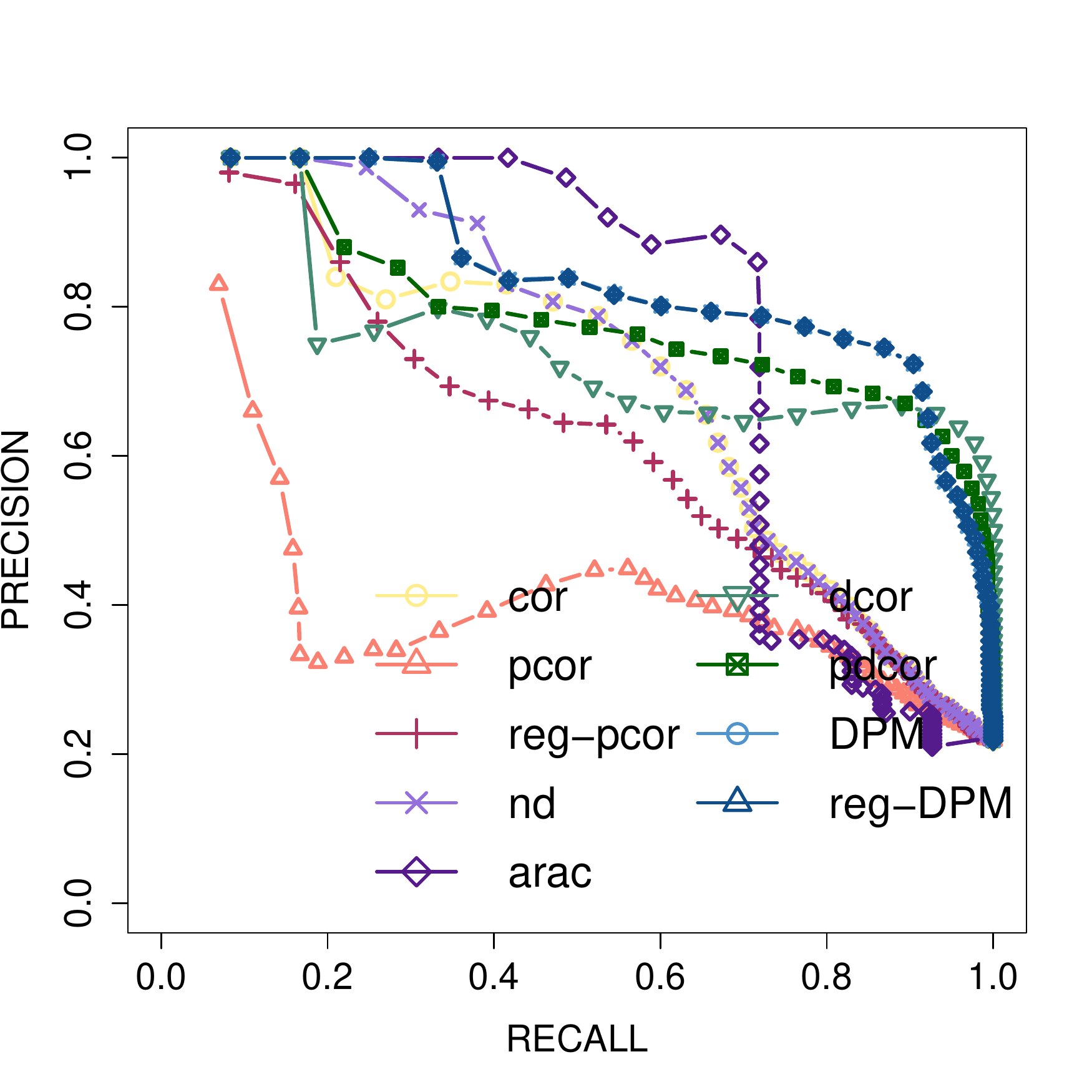}}
\end{tabular}
\caption{\textbf{Performance on nonlinear data.} The upper subplots show the ROC curve, while the lower subplots show the PR curve.}
\label{ngres}
\end{figure}

We proceed to specifically test the ability of DPM to detect nonlinear associations. To this end we simulate 200 samples from the network in Figure \ref{fig:gs}, where the relationships between variables are all nonlinear (see Methods, Simulation of data) and add univariate standard Gaussian noise to each variable. The false positive rate (FPR), recall and precision are computed over 100 replicates. Figure \ref{ngres} shows the average results in the form of ROC and PR curves for different methods. 

In the ROC curves the top group of methods comprises exactly the distance based methods DPM, reg-DPM, dcor, and pdcor. In the PR curves, ARACNE performs best below a recall of $\sim 0.7$. DPM and reg-DPM (overlapping) are below Aracne but maintain their performance also for higher values of recall. Next come pdcor and dcor. Thus, the performance of the distance correlation based methods is better than correlation based methods, in line with their proposed ability to detect non-linear relationships. There is no visible difference between the regularized (reg-DPM) and the naive inversion (DPM), which is in line with the high number of samples for this small number of nodes.

\begin{figure}[htb]
\begin{tabular}{p{.4\textwidth}p{.6\textwidth}}
\centerline{\includegraphics[width=0.42\textwidth]{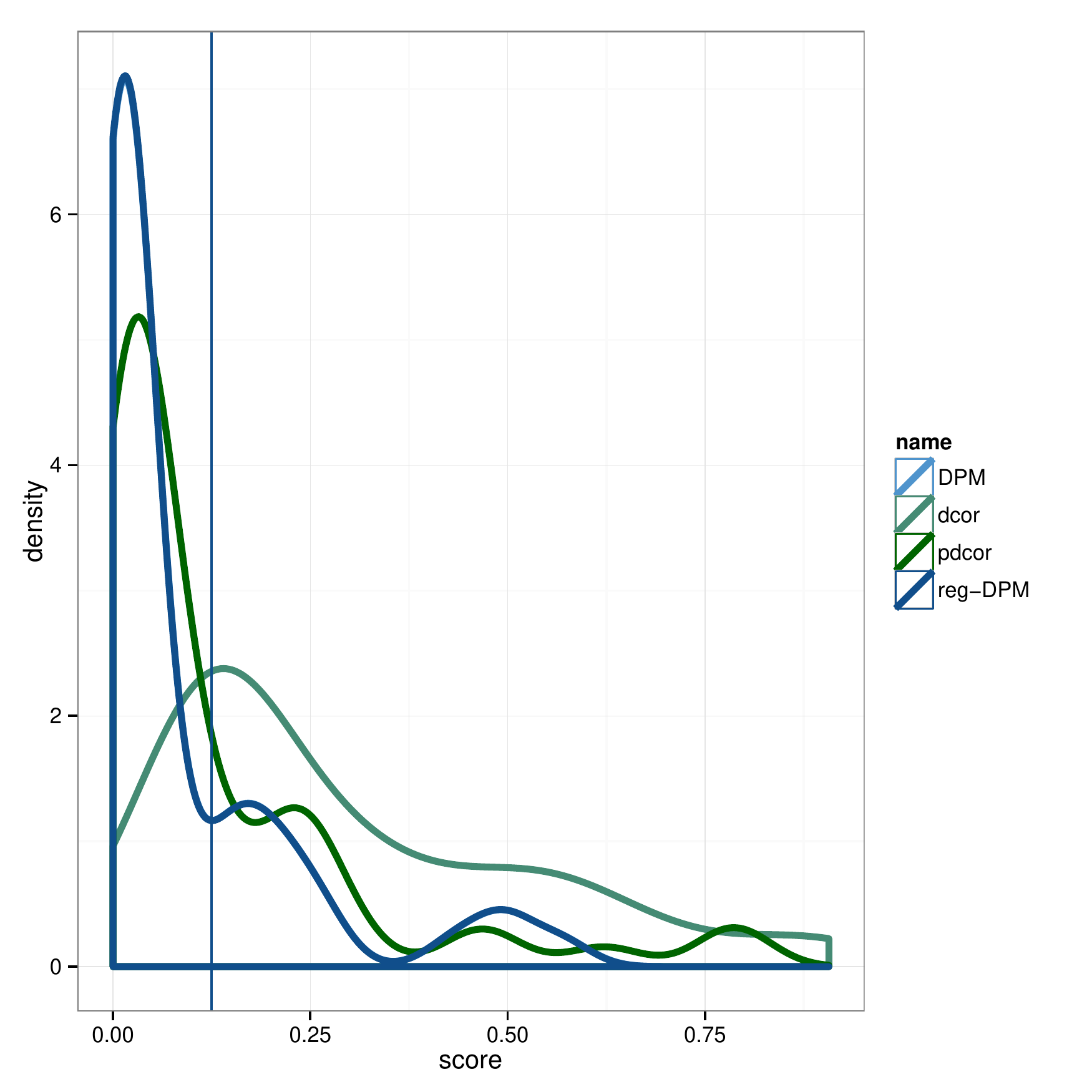}}&
\centerline{\includegraphics[width=0.55\textwidth]{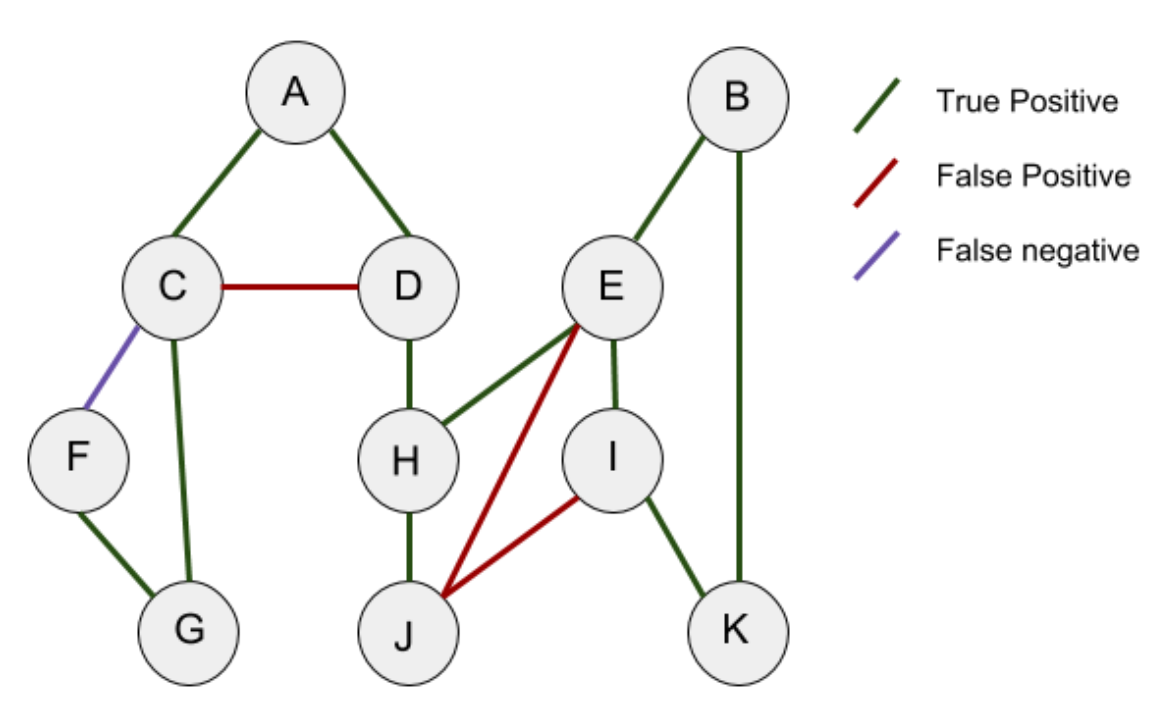}}
\end{tabular}
\caption{\textbf{Threshold selection.} The left subplot shows the density of different scores, while the right subplot shows the network obtained from threshold $0.12$ on reg-DPM score.}
\label{fig:density}
\end{figure}

As pointed out in the Methods Section, ROC and PR serve to compare methods in the presence of a gold standard, but provide little guidance when the method is applied to a novel data-set. We thus study the score distributions of the edges in search for visual clues as to possible cut-offs.
Figure \ref{fig:density} shows the smoothed histograms of the association-scores assigned to the edges by DPM, reg-DPM, dcor, and pdcor (the best methods from the ROC and PR curves). The curves for DPM, reg-DPM, and pdcor are high near 0 and fall off sharply. dcor is not focused on 0 and falls off only slowly.
For illustration purposes we select intuitive thresholds from the histogram (0.12) and depict the respective network in Figure \ref{fig:density}. Edges are colored according to whether they are true/false positives and false negatives are included for easier interpretation.

\subsection{Effect of the number of samples on the performance}
\begin{figure}[htb]
\begin{tabular}{p{.5\textwidth}p{.5\textwidth}}
\centerline{\includegraphics[width=0.55\textwidth]{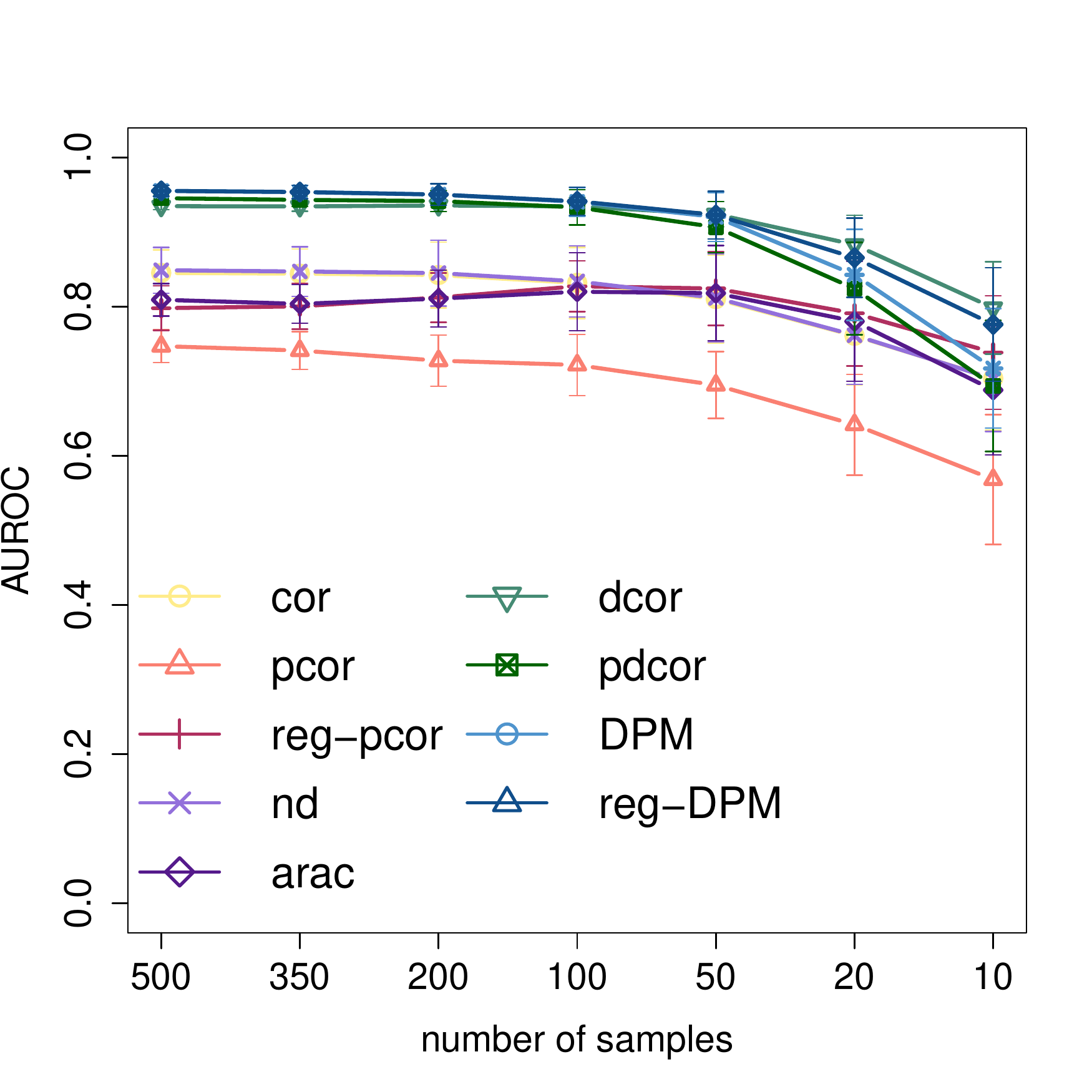}}&
\centerline{\includegraphics[width=0.55\textwidth]{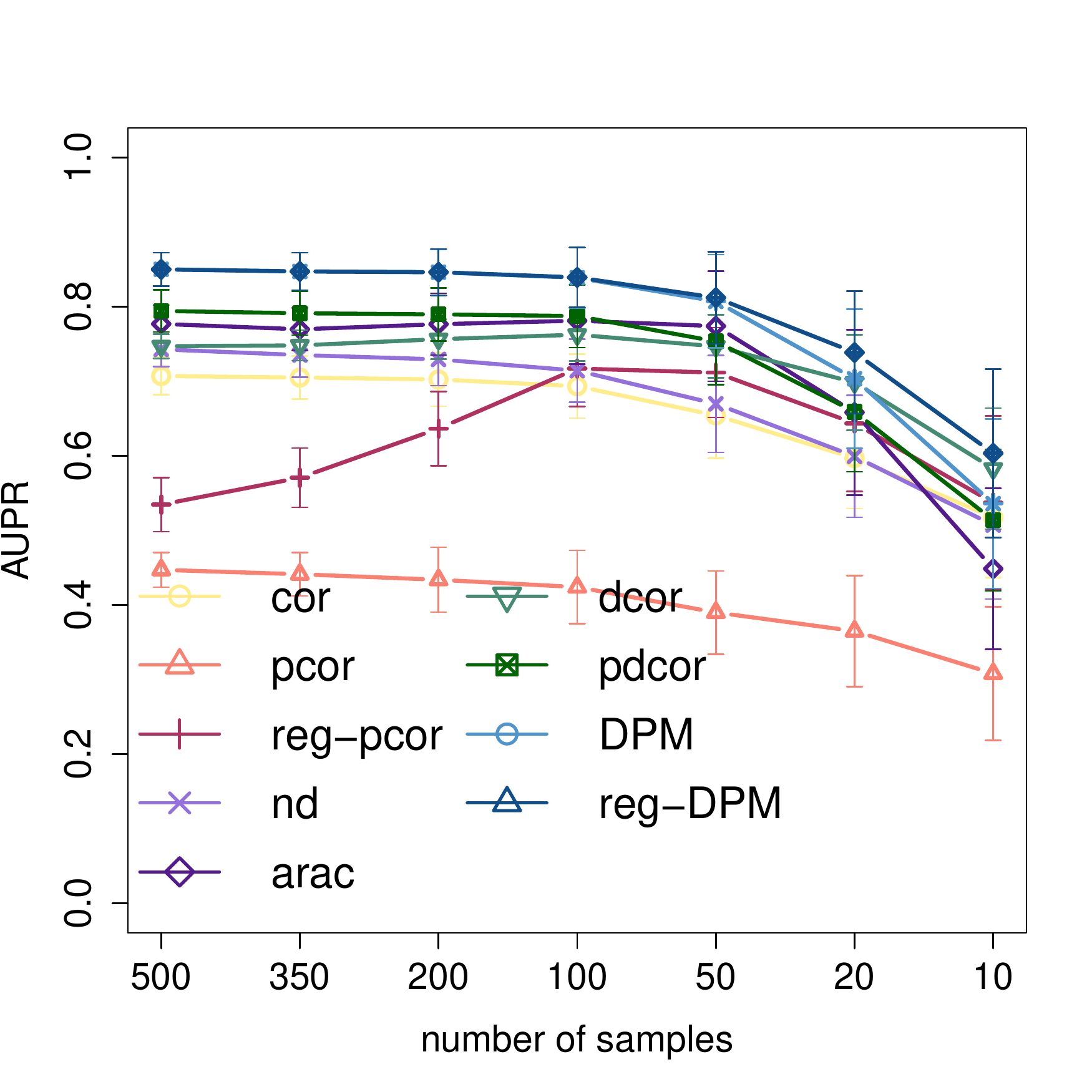}}
\end{tabular}
\caption{\textbf{Effect of the number of samples on the performance.} The left subplot shows the ROC curve, while the right subplot shows the PR curve.}
\label{nsamp}
\end{figure}

Sample size affects the performance of the association scores and our ability to accurately reconstruct networks. We thus assess the effect of the number of samples on the performance of different methods while keeping noise constant. We simulate different number of samples from the non-linear network $GS$ (see Section "simulation of data") and add Gaussian noise $\epsilon \sim N(0,1)$. We repeat this procedure $100$ times and compute the mean and standard deviation for each score for all sample sizes. Figure \ref{nsamp} shows the average AUPRC and AUROC result for different methods and different number of samples. 

In AUROC, for more than 50 samples the distance correlation based methods (DPM, reg-DPM, dcor, pdcor) soundly dominate the other methods (ND, ARACNE, and the correlation based methods). For smaller sample size the top performing methods are distance correlation and reg-DPM, which are above DPM, pdcor and the other methods. In terms of AUPRC, as long as sample size is above 50, DPM and reg-DPM dominate the other methods. With lower sample size, reg-DPM behaves better than DPM and the other methods. 

\subsection{Effect of noise}
\begin{figure}[htb]
\begin{tabular}{p{.5\textwidth}p{.5\textwidth}}
\centerline{\includegraphics[width=0.55\textwidth]{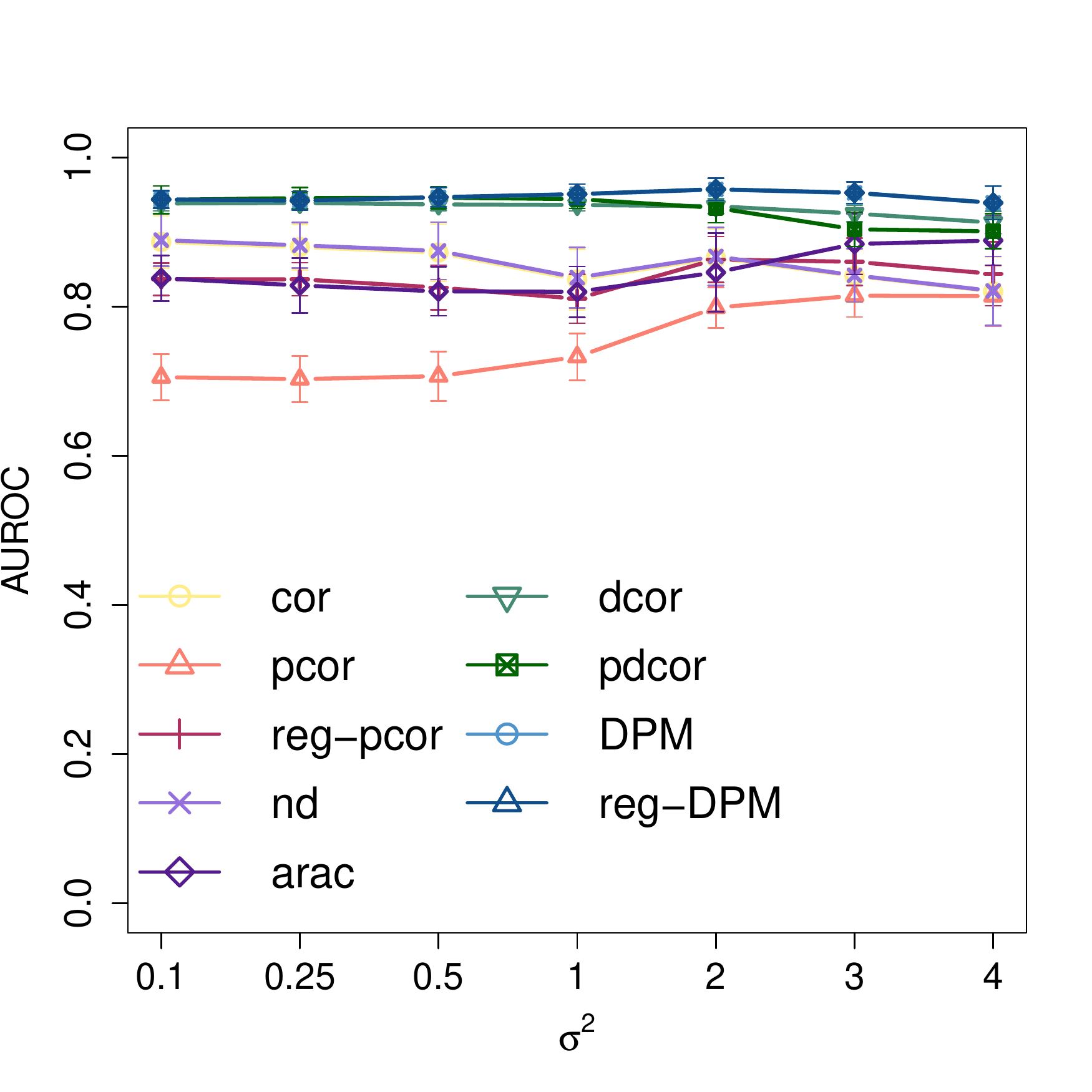}}&
\centerline{\includegraphics[width=0.55\textwidth]{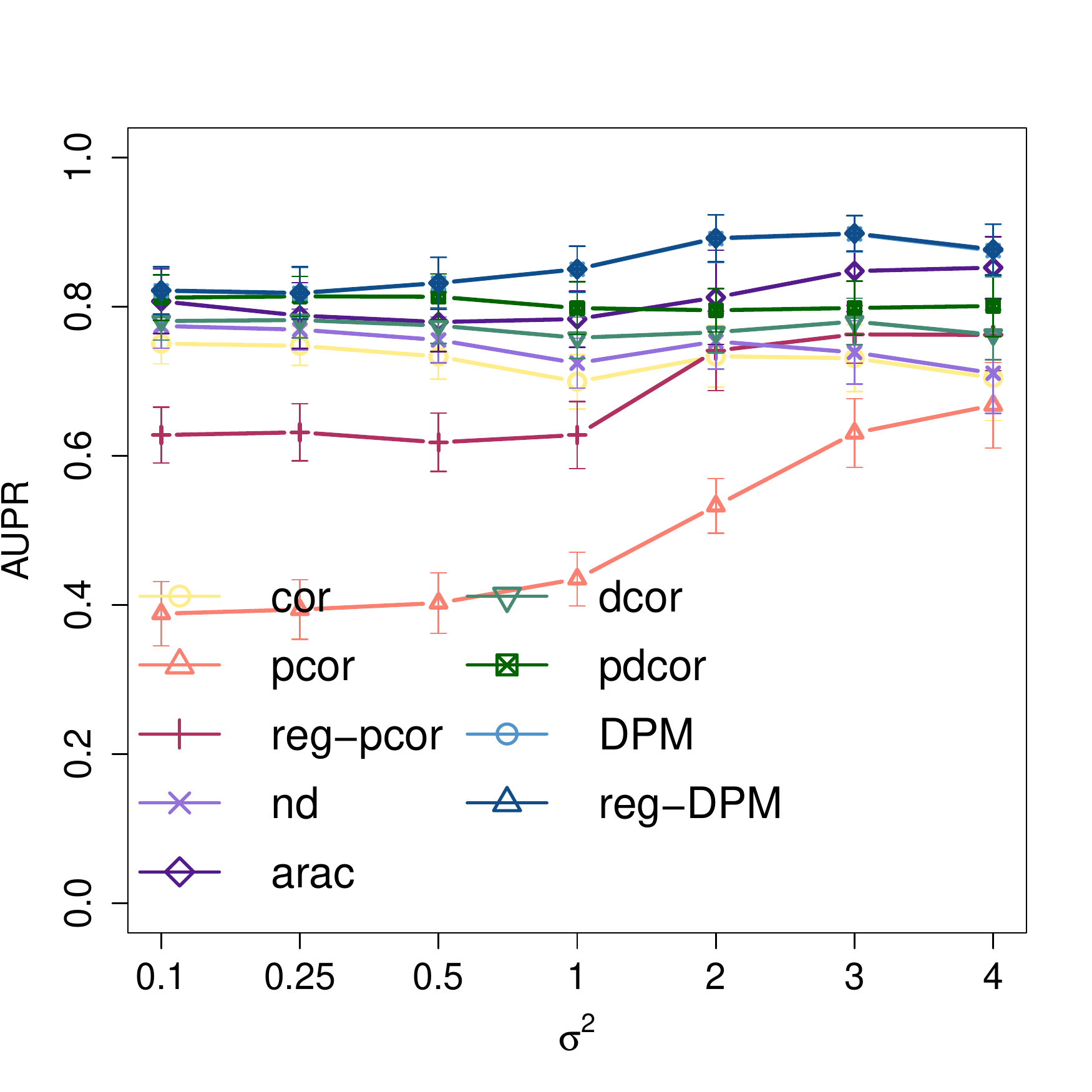}}
\end{tabular}
\caption{\textbf{Effect of the noise on the performance.} The left subplot shows the AUROC results, while the right subplot shows the AUPR results. Simulated datasets comprising 300 samples of nonlinear data each were generated using additive Gaussian noise $\epsilon ~ N(0,\sigma ^2)$}. 
\label{noise}
\end{figure}

We assess the effect of noise on the performance of different methods while keeping the sample size constant. We simulate 300 samples of nonlinear data and add Gaussian noise $\epsilon ~ N(0,\sigma ^2)$ with $\sigma ^2$ going from 0.1 to 4. We repeat this procedure $100$ times and compute the mean and standard deviation for each score at all noise level. Figure \ref{noise} shows the effect of noise on the AUPRC and the AUROC for different methods. 

As indicated by the average AUC the performance of all methods decreases with increasing noise. Both AUROC and AUPRC graphs indicate that all four distance correlation based methods perform better than the other methods. Among the distance correlation based methods, the DPM and reg-DPM perform well above the others especially in terms of AUPRC. 

\subsection{Application to gene regulatory networks (DREAM)}

\begin{figure*}[ht]
 \begin{tabular}{p{.45\textwidth}p{.45\textwidth}}
 \centerline{\includegraphics[width=.4\textwidth]{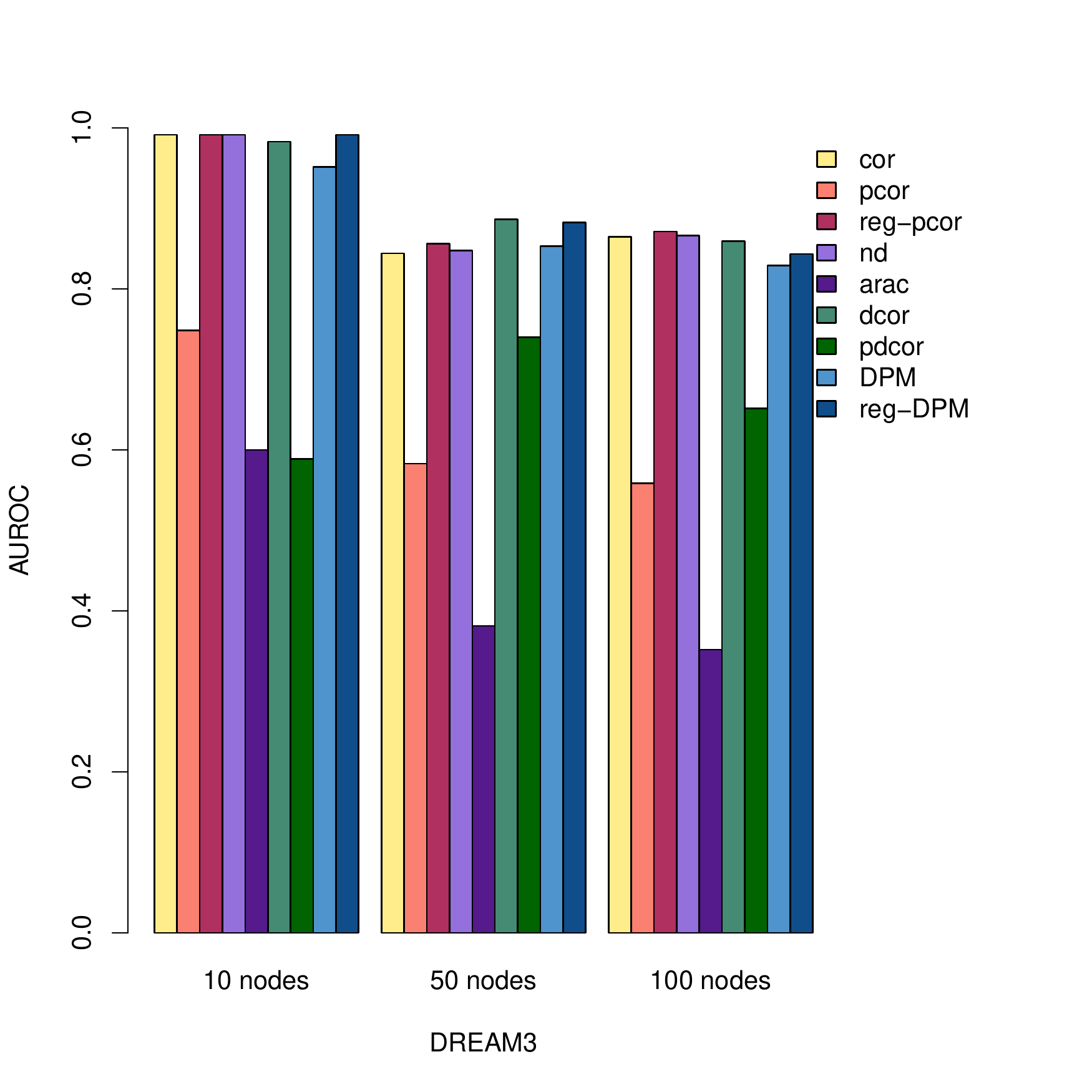}}&
 \centerline{\includegraphics[width=.4\textwidth]{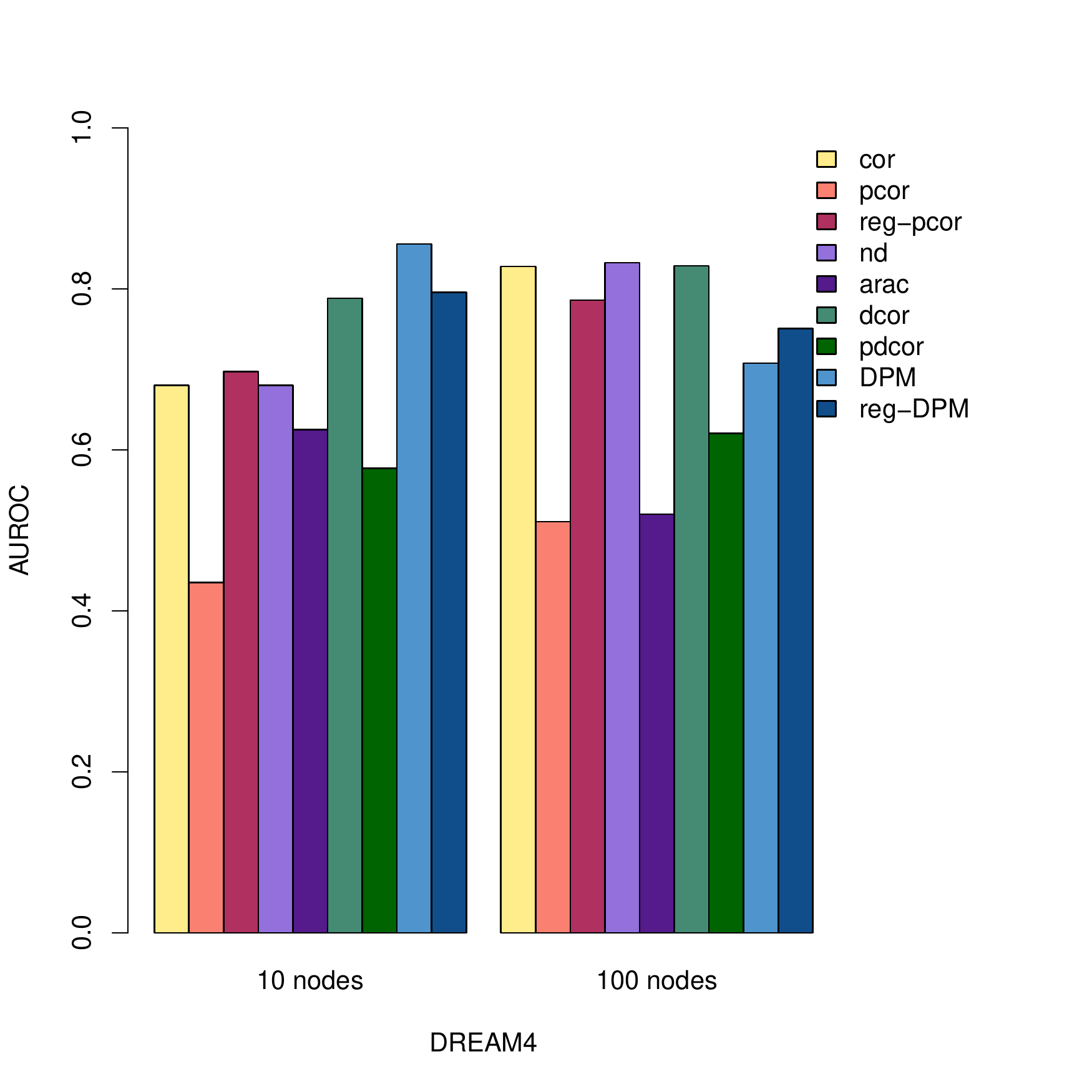}}\\
 \centerline{\includegraphics[width=.4\textwidth]{D4aurocbarplot.pdf}}&
 \centerline{\includegraphics[width=.4\textwidth]{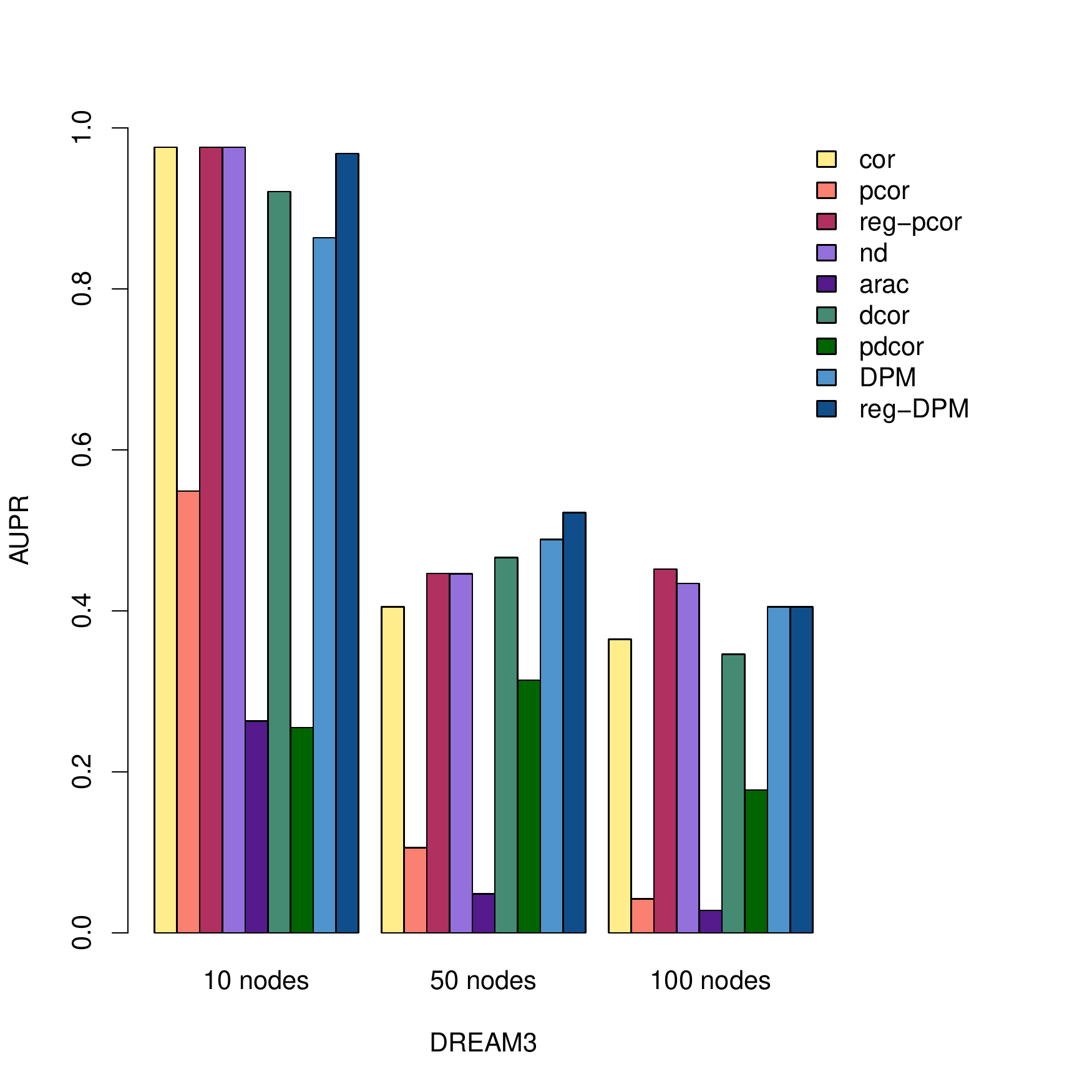}}\\
 \end{tabular} 
 \caption{\textbf{Performance on DREAM challenge data.} The upper subplots show the AUROC results, while the lower subplots show the AUPR results. }
  \label{DREAM}
  \end{figure*}

In this section, we compare the performance of the different methods in particular for reconstruction of gene regulatory networks by using data from the DREAM challenge (see Methods).
Figure \ref{DREAM} shows the AUROC and AUPR results for data from DREAM3 and 
4.

On the simulated DREAM3 data one can identify a group of good methods comprising cor, (reg-)DPM, reg-pcor, ND, and dcor. From amongst the distance based methods, pdcor in this setting performs worse. Only for the small network with 10 nodes do these methods achieve an AUROC of almost 1, while for the larger networks it is around 0.8. For all methods, the AUPRC for the networks with 50 or 100 nodes is low. 

On the DREAM4 simulated data, in the small network DPM and reg-DPM are better than dcor in terms of AUROC. On the larger network, the top methods are cor, ND, dcor, reg-pcor, followed by (reg-)DPM. In terms or AUPRC, in the small network the top methods are the same as for AUROC. For the larger networks, even the best method, reg-pcor, achieves an AUROC of only $ \sim 0.25$.

\section{Discussion}

In terms of methods, the Distance Precision Matrix introduced here constitutes a generic approach to network reconstruction. Drawing on distance correlation as its substrate, it is not limited to the detection of linear relationships but capable of recognizing non-linear associations. 

The simulation of a network from non-linear data showed that the distance correlation based methods are best suited to reconstruct those relationships successfully. Considering both AUROC and AUPRC DPM and reg-DPM show consistently good performance on non-linear data. For the simulated data from the DREAM3 and DREAM4 challenge it is difficult to find a consistent trend as to the best methods. There appear to be cases where correlation based methods perform well, although generally the distance correlation based methods are highly competitive.

The network aspect requires a remedy against the transitive carry-over of signal via several nodes. This has traditionally been the precision matrix, which concentrates the signal on the direct interactions. Since distance correlation maps the original variables into a high dimensional space, where it again performs linear operations, the combination of both worlds uses the variance-covariance matrix from the high-dimensional space and inverts it. This, in turn, should be done using a regularization method, in particular when the number of available samples is low.
This is certainly the case in genetics. Here, as well as in classification, this is known as the p$>>$n problem, where n corresponds to the number of samples. Clearly, we cannot "solve" this problem. It is a data problem and not a methods problem. Our simulations show the behavior of several methods in dependence of the number of samples. Especially on the difficult problems where the number of samples is low, the strength of reg-DPM becomes apparent. 
 
On the other hand, the simulations with a large number of samples allow us to inquire about consistency. 
While some methods are far from being consistent in their reconstructions, other methods once they are given a certain number of samples do get close to the correct network. This is encouraging, but does not yet solve all network reconstruction problems in biology, because there sample sizes are still low and the regularization is an essential remedy.

It is important to point out that the original inventors of distance correlation, Szekely and coworkers \cite{dcor} also propose a form of partial distance correlation \cite{pdcor}, which is more sophisticated than the our proposition. They have good mathematical reasons for their definition, and yet our simulations indicate that the simple approach introduced here performs better in practice. Especially on the DREAM data, their proposed methods gave less satisfactory results. Why this is the case would be a topic for further study. One might speculate, that if only their definition would allow for robustification by a regularization method, it would again work better. However, it is non-obvious how to achieve this. This is why we have opted for the simple definition of partial distance correlation and gained the ability to compute the Distance Precision Matrix by a regularization method.

Although the Distance Precision Matrix is intuitive and simple, we cannot claim that a zero-entry in it is equivalent to conditional independence of the respective variables. In fact, \cite{pdcor} contains a (contrived)  example where this cannot possibly be the case. Our simulations and examples, however, support the assumption, that the DPM method is useful even without this "holy grail" assertion.

\bibliographystyle{siam}
\bibliography{dpm}

\end{document}